# Topological transitions in superconductor nanomembranes in a magnetic field with submicron inhomogeneity under a strong transport current

R. O. Rezaev[1,2], E. I. Smirnova[1], O. G. Schmidt[1,3-5], V. M. Fomin[1,6*]

[1]*Institute for Integrative Nanosciences, Leibniz IFW Dresden, Helmholtzstraße 20, D-01069 Dresden, Germany*

[2]*Tomsk Polytechnic University, Lenin av. 30, 634050 Tomsk, Russia*

[3]*Material Systems for Nanoelectronics, TU Chemnitz, Reichenhainer Straße 70, D-09107 Chemnitz, Germany*

[4]*Research Center for Materials, Architectures and Integration of Nanomembranes (MAIN), TU Chemnitz, Rosenberg Straße 6, D-09126 Chemnitz, Germany*

[5]*Nanophysics, Faculty of Physics, TU Dresden, Nöthnitzer Straße 61, D-01187 Dresden, Germany*

[6]*National Research Nuclear University MEPhI, Kashirskoe shosse 31, 115409 Moscow, Russia*

## Abstract

**Under a strong transport current, the induced voltage in superconductor nanomembranes in a magnetic field with submicron inhomogeneity shows a pulse on a certain interval of the magnetic field. It is a manifestation of a wide phase-slip domain. The topological transition accompanying the phase-slip effect consists in the occurrence of two new loops of weak superconducting currents connecting two regions of superconducting screening currents, which are disconnected in case of vortex-chain dynamics. In the middle of the phase-slip domain, rapid dynamics of the superconducting order parameter consists in decoupling of a spontaneously nucleated vortex-antivortex pair, subsequent motion of a vortex and an antivortex in the opposite directions followed by their annihilation with an antivortex and a vortex from the adjacent pairs. The submicron-scale inhomogeneity of the magnetic field can be achieved through a direct patterning of the magnetic field applied to a planar membrane or using advanced nanostructuring, such as roll-up technology, focused ion-beam deposition or coating carbon nanotubes by superconducting materials. If the applied magnetic field is**

**orthogonal to the axis of a microtube, which carries transport current in the azimuthal direction, the phase-slip regime is characterized by the vortex-antivortex lifetime of $10^{-15}$ s versus $10^{-12}$ s for disconnected vortex dynamics in the half-tubes. The phase-slip dynamics determines the voltage-magnetic field and voltage-current characteristics in nanoarchitectures with multiple disconnected loops of superconducting screening currents.**

Topological defects in a superconductor, where the order parameter locally vanishes and its phase has no definite value, lead to the emergence of a finite resistance: "the superconductor is no longer "superconducting" in a practical sense: it offers resistivity to the current!" [1]. The most well-known topological defects are vortices (antivortices) in a 3D superconductor, where the superconducting order parameter decreases from a certain value in the region far from a defect to zero on a line called vortex core [2]. The phase of the order parameter is not defined on the vortex core. When encircling a vortex core, the phase of the order parameter changes by $\pm 2\pi$ for ordinary vortices/antivortices or by an integer multiple of $\pm 2\pi n$ ($n=2,3…$) for giant vortices/antivortices.

Under certain circumstances, topological defects occur in confined superconductor structures. The total phase advance by a multiple of $2\pi$ around a topological defect is called *phase slip*. The concept of the phase slippage in the resistive state of the narrow (quasi-1D) superconductor filaments was introduced in [3]. At the phase slip event, both Re $\psi$ and Im $\psi$ vanish at a point along the filament. The analysis in [3] was based on the thermal-fluctuation-dominated regime of the occurrence of phase slips near the critical temperature. Fluctuation-driven phase slips may be considered as a sequence of the events of local vanishing and local reappearance of the order parameter. Following [4], let us assume, that in a quasi-1D superconductor carrying transport current just below the critical temperature, there occurs a weak link (a domain of weakened superconductivity). In case if the transport current is slightly increased, the superconducting state vanishes in the weak-link domain. In the next instance, the weak-link domain reveals normal conductivity mediated by unpaired electrons, whose drift velocity is reduced by inelastic scattering. As long as there are unpaired electrons below the critical temperature with a drift velocity smaller than the critical velocity of the condensate, those electrons condense into Cooper pairs, which are accelerated to carry the current as a supercurrent, and the process repeats again.

Later, other mechanisms of phase slips were developed. In particular, inhomogeneity of a superconductor filament [5] makes its critical current inhomogeneous, so that an increase in the transport current leads to a local collapse of the order parameter. The entire current must be then carried as the normal current. This allows superconductivity to reappear, and the cycle to repeat. For low temperatures below 0.2 K, the behavior of a 1D superconductor was shown to be dominated by quantum-mechanical tunneling through the free-energy barrier: *quantum phase slips* [6]. An experimental evidence for a quantum phase transition driven by the quantum fluctuations in a 2D superconductor has been recently obtained through magneto-transport measurements [7].

A further insight in the phase-slip phenomenon was gained for 2D and quasi-2D superconductors. For nanowires of 100 nm width (nanostripes), it was shown [8-9] that at relatively low current densities the one-dimensional Langer-Ambegaokar-McCumber-Halperin (LAMH) mechanism [3,10] based on thermally induced phase slip centers dominates over the two-dimensional mechanism related to unbinding of vortex – antivortex pairs below the Berezinskii-Kosterlitz-Thouless (BKT) transition in 2D [11-13]. A BKT-type crossover was detected in a trapped quantum degenerate gas of rubidium atoms [14]. At temperatures above the BKT critical temperature up until the critical temperature, thermal fluctuations are sufficient to unbind vortex-antivortex pairs, while below the BKT critical temperature all vortices are tightly bound into vortex-antivortex pairs [15]. The effect of the magnetic-field induced BKT scenario was observed in a 2D spin-dimer system by using a multilayer magnet [16]. Another example of a phase transition in superconducting vortex systems is a liquid – solid transtion experimentally found in 2D geometry [17].

At weaker transport currents, dissipation below the critical temperature is attributed to thermally activated motion of the order parameter over the free-energy barriers, which separate metastable states. At stronger transport currents, the free energy required for unbinding a vortex-antivortex pair is reduced, and the corresponding thermally activated contribution to the resistivity is expected to be dominant. In this regime, the resistivity due to vortex flow is substantially affected by electron heating. In an external magnetic field, the situation changes drastically. Even in a weak magnetic field, the contribution of moving vortices dominates over the thermally activated phase slip mechanism in the resistivity in superconducting wires of submicron width [8].

In novel superconductor nanostructured microarchitectures, e.g., open microtubes [18-19] and microcoils [20-22] in a magnetic field orthogonal to their axes, distribution of the order parameter over the surface may be highly inhomogeneous. This opens up new possibilities for manifestations of phase slips as compared to quasi-1D filaments and quasi-2D stripes. In the nanohelices as small as 100 nm in diameter and aspect ratio up to 65, fingerprints of vortex and phase-slip patterns have been experimentally identified and supported by numerical simulations based on the time-dependent Ginzburg−Landau equation [22]. In the present paper, we will show that a transition between different patterns of vortices is accompanied by occurrence of a weak link at the side of a microtube, which is opposite to the slit, giving rise to novel phase slip events.

Quasi-2D superconducting nanoarchitectures provide a unique possibility to study the interplay of multiple disconnected regions of superconducting screening currents (SSCs), which are induced by an external magnetic field. The dynamics of topological defects (superconducting vortices and phase slips) and the topological transitions between different configurations of SSCs manifest themselves through voltage-current or voltage-magnetic field characteristics.

The simplest example of multiple disconnected loops of SSC is the two loops of SSC (Fig. 1**c**), which can be realized experimentally in an open superconducting tube placed in a homogeneous magnetic field orthogonal to the tube's axis (Fig. 1**a**). The 2D-cylindrical structure is convenient for the numerical study of the order-parameter dynamics, since the mathematical model can be formulated in a way equivalent to the planar structure in an inhomogeneous magnetic field. The tube has a cut in the paraxial direction. Two electrodes are attached to both edges of the slit (see Fig. 1**a**), in order to support an azimuthal transport current. The width of the slit is supposed to be much smaller than the circumference length $2\pi R$ and the electrodes extend along the entire length of the edges. The whole system is placed in the magnetic field $\mathbf{B} = B\mathbf{e}_\mathbf{z}$, which induces SSC (Meissner currents) circulating in each half-tube. The superconducting state is described by the time-dependent Ginzburg-Landau (TDGL) equation for the complex-valued order parameter $\psi$ coupled with the Poisson's equation (see Methods).

The SSC pattern for the planar structure is represented in Fig. 2**a**-2**f**, where the transport current is applied in the $\theta$-direction. The superconducting current after subtraction of the transport current is presented in Figs. 2**c**, **f**. The solid lines are the current lines. Each vortex is a topoligical defect for the phase of the order parameter, while the phase-slip appearance is accompanied by the topological transition of SSCs. For the planar structure, the spacio-temporal analysis of the order parameter reveals a topological transition from *one* loop of SSCs (Fig. 2**c**) to *three* loops (Fig. 2**f**) when the phase slip appears. However, no topological changes are detected in the vortex dynamics, when the phase-slip has appeared. The phenomenon of flux-flow instability that governs the occurrence of localized regions with suppressed superconductivity was described theoretically in a disordered thin superconducting strip in a homogeneous magnetic field [23]. In an open nanotube, we observe that a region of the fast-moving vortices (shown in Fig. 2**g, h**) near an edge transforms into a phase-slip region, which propagates to the opposite edge of the structure. Two loops of SSC are demonstrated in Fig. 2**i**.

The phase-slip induced topological transition of SSC is shown in Fig. 3, where *two* loops of SSC transform into *four* loops. As distinct from the planar structure, this transition is associated with the topological changes in the vortex dynamics – ***vortex-antivortex pairs*** start to be generated at a frequency, which is three orders of magnitude higher than that for vortex chains. The manifestation of topological nature of vortices is shown in Fig. 3**d**, which depicts the phase of the order parameter. When encircling each vortex/antivortex core in a specific direction (for example, counter-clockwise) the phase changes by $+2\pi$ for a vortex and by $-2\pi$ for an antivortex (see the black and white circles in Fig. 3**e**). The spacio-temporal structure of the phase slip is thus topologically different from the trivial topology of the initial two loops of SSC. The vortex-antivortex pairs generation has a striking similarity with the BKT-transtion obsreved in 2D systems (XY-models). Our simulation shows that when the magnetic field increases, the superconducting state without a phase slip reenters the structure with a chain of moving vortcies, which are then trasformed into channels of fast moving vortices (Fig. 3**c, f**). Such channels are supposed to be observed experimentally [24]. As a result of the sequence of transitions superconducting vortex state–phase-slip state–superconducting vortex state, a voltage pulse (Fig. 3**e**) is expected to be observed in the open tube, while in the planar structure, a monotonic growth of the voltage as a function of the magnetic field occurs.

The obtained dynamics of the phase of the order parameter reflects the dynamics of kinematic vortex–antivortex pairs, first theoretically proposed in Ref. [25] and experimentally detected in Ref. [26] in planar microstructures. Kinematic vortex-antivortex pairs were shown [26] to have velocities higher than the Abrikosov vortex velocities. In Refs. [25, 27], kinematic vortex–antivortex pairs were numerically described under strong transport current (close to the critical value) at zero and weak magnetic fields. In the tubular structure (Fig. 1**a**) multiple kinematic vortex-antivortex pairs nucleate under strong transport current at the regions with zero and weak normal-to-surface component of the magnetic field. Remarkably, the green loops of SSC in Fig. 3**b** connect front and rear half-tubes, thus introducing a new topology of the superconducting current. A topological transition from two disconnected regions of SSCs (red) to four (red and green) and then back to two (red) such regions occurs with increasing the magnetic field for certain intervals of values of the transport current. Under these conditions in open tubes, the nucleation locus connects **both halfs** of the tube (Fig. 3**b**), as distinct from the previously known cases when kinematic vortex-antivortex pairs do not occur and the both halfs of the tube are strictly disconnected (Fig. 1**c**). Generation of the vortex-antivortex pairs marked with white and pink circles in Fig. 3**d** results from (i) their unbinding due to the high transport current and (ii) motion due to the Magnus force caused by the magnetic field. Nucleation and separation of vortex-antivortex pairs at the side of the microtube, which is opposite to the slit, are followed by their motion till (i) their denuncleation at the sides of the tube or (ii) annihilation of a vortex from the pair with an antivortex from a neighboring pair (resp. an antivortex from the pair with a vortex from another neighboring pair), when there exist two or more vortex-antivortex pairs. The relatively fast motion of the vortices and antivortices on the side of the microtube, which is opposite to the slit, leads to an apparent picture of an extended static phase slip (Fig. 3**a**). As follows from the comparison with the calculated vortex dynamics in an open tube, the ratio of velocities for kinematic vortex-antivortex pairs and Abrikosov vortices is as large as about 2 orders of magnitude, what is close to that obtained for a Sn planar film [26].

The voltage generated in DC transport represents different order-parameter states: the pure superconducting state, the mixed state (vortices and the superconducting state), the phase-slips states and the normal state. A switching between those states corresponds to jumps in voltage as a function of the current or the magnetic field. The voltage pulse shown in Fig. 3 is calculated for the currents close to the critical ones (i.e. higher than $0.5J_C$). Within these values of parameters, the state of the order parameter is estimated to be very sensitive to the temperature. In Fig. 4, we represent the calculated voltage response for Sn which has different values of the coherence length and the magnetic field penetration depth as compared to Nb, under different temperature conditions. The order parameter state in Nb is more sensitive to temperature variations in comparison with Sn (as follows from the comparison of **SI** Fig. S2 and Fig. S3).The temperature range, where the phase-slip regime exists, is wider for the Sn microtube than for the Nb one: the voltage pulse in Nb occurs for temperature varying within 1 per cent of $T_c$, while for Sn this temperature range is increased up to ~7 per cent of $T_c$. The voltage derivative as a function of the magnetic field demonstrates a relatively small peak for the case when vortices with a characteristic dimension ~$4\xi$ each form a chain with a length just equal to the length $L$ of the tube.

A nontrivial superconducting currents topology induces the phase-slip dynamics, which determine the voltage-magnetic field and voltage-current characteristics in nanoarchitectures with multiple loops of SSC. The spacio-temporal structure of the phase slips in an open tube and a planar structure is unveiled. The crucial difference between them lies in the vortex-antivortex pair generation. The non-monotonous voltage-magnetic field and voltage-current characteristics imply a possibility to efficiently tailor the superconducting properties of nanostructured materials by inducing a nontrivial topology of SSCs.

## Acknowledgement

Authors thank A. Bezryadin, J. Lorenzana, D. Roditchev, V. A. Shklovskij, H. Suderow, F. Tafuri, R. Tidecks, V. M. Vinokour, A. D. Zaikin, and E. Zeldov for fruitful discussions. Authors are grateful to DFG for support under the project # FO 956/5-1 and to ZIH TU Dresden for providing its facilities for high throughput calculations. R.O.R. thanks RFBR and Tomsk Region for support through the research project # 19-41-700004.

## Author contributions

V. M. F. led the project and conceived the investigations. R. O. R. and E. I. S. performed numerical simulations. O. G. S. and V. M. F. developed key conceptual ingredients for the physical interpretation. V. M. F. and R. O. R. wrote the manuscript. All authors contributed to discussions about the numerical modeling, the analysis of the obtained results and the preparation of the final manuscript.

## Competing interests

The authors declare no competing interests.

## References

1. N. B. Kopnin, *Theory of Nonequilibrium Superconductivity*, Clarendon Press, Oxford, 2001.
2. M. Tinkham, *Introduction to Superconductivity*, McGraw-Hill, New York, 1996.
3. J. S. Langer and V. Ambegaokar, Intrinsic Resistive Transition in Narrow Superconducting Channels, Phys. Rev. 164, 498-510 (1967).
4. R. Tidecks, *Current-Induced Nonequilibrium Phenomena in Quasi-One-Dimensional Superconductors*, Springer Tracts in Modem Physics, vol. 121, Springer, Berlin-Heidelberg, 1990, 341 pp.
5. J. Skocpol, M. R. Beasley and M. Tinkham, Phase-Slip Centers and Nonequilibrium Processes in Superconducting Tin Microbridges, J. Low Temp. Physics 16, 145-167 (1974).
6. N. Giordano, Evidence for Macroscopic Quantum Tunneling in One-Dimensional Superconductors, Phys. Rev. Lett. 61, 2137-2140 (1988).
7. Y. Saito, T. Nojima, and Y. Iwasa, Quantum phase transitions in highly crystalline two-dimensional superconductors, Nature Communications 9, 778, 1-7 (2018).
8. M. Bell, N. Kaurova, A. Divochiy, G. Gol'tsman, J. Bird, A. Sergeev, and A. Verevkin, On the Nature of Resistive Transition in Disordered Superconducting Nanowires, IEEE Trans. on Appl. Sup. 17, 267-270 (2007).
9. M. Bell, A. Sergeev, V. Mitin, J. Bird, A. Verevkin, G. Gol'tsman, One-dimensional resistive states in quasi-two-dimensional superconductors: Experiment and theory Phys. Rev. B 76, 094521, 1-5 (2007).

10. D. E. McCumber and B. I. Halperin, Time Scale of Intrinsic Resistive Fluctuations in Thin Superconducting Wires, Phys. Rev. B 1, 1054-1070 (1970).
11. V. S. Berezinskii, Destruction of Long-range Order in One-dimensional and Two-dimensional Systems having a Continuous Symmetry Group I. Classical Systems, Sov. Phys. JETP 32, 493-500 (1971).
12. V. S. Berezinskii, Destruction of Long-range Order in One-dimensional and Two-dimensional Systems having a Continuous Symmetry Group II. Quantum Systems, Sov. Phys. JETP 34, 610-616 (1971).
13. J. M. Kosterlitz and D. J. Thouless, Ordering, metastability and phase transitions in two-dimensional systems, J. Phys. C 6, 1181-1203 (1973).
14. Z. Hadzibabic, P. Krüger, M. Cheneau, B. Battelier, and J. Dalibard, Berezinskii-Kosterlitz-Thouless crossover in a trapped atomic gas, Nature 441, 1118-1121 (2006).
15. B. I. Halperin and D. R. Nelson, Resistive Transition in Superconducting Films, Journal of Low Temperature Physics 36, 599-616 (1979).
16. U. Tutsch, B. Wolf, S. Wessel, L. Postulka, Y. Tsui, H. O. Jeschke, I. Opahle, T. Saha-Dasgupta, R. Valenti, A. Brühl, K. Remović-Langer, T. Kretz, H.-W. Lerner, M. Wagner, and M. Lang, Evidence of a field-induced Berezinskii-Kosterlitz-Thouless scenario in a two-dimensional spin-dimer system, Nature Communications 5, 5169, 1-9 (2014).
17. B. Chen, W.P. Halperin, P. Guptasarma, D.G. Hinks, V.F. Mitrovic, A.P. Reyes, and P.L. Kuhns, Two-dimensional vortices in superconductors, Nature Physics 3, 239-242 (2007).
18. V. M. Fomin, R. O. Rezaev, and O. G. Schmidt, Tunable generation of correlated vortices in open superconductor tubes, Nano Lett. 12, 1282-1287 (2012).
19. R. O. Rezaev, E. A. Posenitskiy, E. I. Smirnova, E. A. Levchenko, O. G. Schmidt and V. M. Fomin, Voltage Induced By Superconducting Vortices In Open Nanostructured Microtubes, Phys. Stat. Sol. RRL 13, 1-12 (2019).
20. V. M. Fomin, R. O. Rezaev, E. A. Levchenko, D. Grimm and O. G. Schmidt, Superconducting nanostructured microhelices, Journal of Physics: Cond. Mat. 29, 395301, 1-9 (2017).
21. S. Lösch, A. Alfonsov, O. V. Dobrovolskiy, R. Keil, V. Engemaier, S. Baunack, G. Li, O. G. Schmidt, and D. Bürger, Microwave Radiation Detection with an Ultra-Thin Free-Standing Superconducting Niobium Nanohelix, ASC Nano 13, 2948–2955 (2019).

22. R. Córdoba, D. Mailly, R. O. Rezaev, E. I. Smirnova, O. G. Schmidt, V. M. Fomin, U. Zeitler, I. Guillamón, H. Suderow, J. M. De Teresa, Three-dimensional superconducting nano-helices grown by He+-focused-ion-beam direct writing, Nano Lett. 19, 8597-8604 (2019).
23. D. Yu. Vodolazov, Flux-flow instability in a strongly disordered superconducting strip with an edge barrier for vortex entry, Supercond. Sci. Technol. 32, 115013, 1-7 (2019).
24. L. Embon, Y. Anahory, Ž.L. Jelić, E. O. Lachman, Y. Myasoedov, M. E. Huber, G. P. Mikitik, A. V. Silhanek, M. V. Milošević, A. Gurevich, E. Zeldov, Imaging of super-fast dynamics and flow instabilities of superconducting vortices, Nature Communications 8, 85, 1-10 (2017).
25. A. Andronov, I. Gordion, V. Kurin, I. Nefedov and I. Shereshevsky, Kinematic vortices and phase slip lines in the dynamics of the resistive state of narrow superconductive thin film channels, Physica C 213, 193-199 (1993)
26. A. G. Sivakov, A. M. Glukhov, A. N. Omelyanhchouk, Y. Koval, P. Müller, and A. V. Ustinov, Josephson behavior of phase-slip lines in wide superconducting strips, Phys. Rev. Lett., 91, 267001, 1-4 (2003).
27. G. R. Berdiyorov, M. V. Milosevic, and F. M. Peeters, Kinematic vortex-antivortex lines in strongly driven superconducting stripes, Phys. Rev. B 79, 184506, 1-8 (2009).
28. R. Kato, Y. Enomoto, and S. Maekawa, Effects on the surface boundary on the magnetization process in type-II superconductors, Phys. Rev. B 47, 8016-8024 (1993).
29. R. D. Parks, *Superconductivity*, vol. 1, Marcel Dekker, 1969.
30. Y. Nakamura, T. Zhao, J. Xi, W. Shi, D. Wang, and Z. Shuai, Intrinsic charge transport in stanene: roles of bucklings and electron-phonon couplings, Adv. Electron. Mater. 3, 1700143,1-9 (2017).
31. S. Tanuma, C. J. Powell, and D. R. Penn, Calculations of electron inelastic mean free paths. IX. Data for 41 elemental solids over the 50 eV to 30 keV range, Surface and Interface Analysis 43, 689-713 (2011).

# Methods

The system is placed in the applied magnetic field $\mathbf{B} = B\mathbf{e}_z$, which induces SSC (Meissner currents) circulating in each half-tube. The superconducting state is described by the time-dependent Ginzburg-Landau (TDGL) equation for the complex-valued order parameter $\psi$ in the dimensionless form (see Tables 1, 2):

$$\frac{\partial \psi}{\partial t} = -\left(\frac{1}{i\kappa}\nabla - \mathbf{A}\right)^2 \psi + (1 - |\psi|^2)\psi - i\kappa\varphi\psi, \tag{1}$$

where $\mathbf{A}$ is the vector potential; $\varphi$ the scalar potential and $\kappa = \lambda/\xi$ the Ginzburg–Landau parameter with the London penetration depth $\lambda$ and the coherence length $\xi$. Boundary conditions follow from the absence of the normal component of the superconducting current at the free boundaries of the superconductor:

$$\left(\mathbf{n}, \frac{1}{i\kappa}\nabla - \mathbf{A}\right)\psi\bigg|_{\partial D_s} = 0; \quad \left(\mathbf{n}, \frac{1}{i\kappa}\nabla - \mathbf{A}\right)\psi\bigg|_{\partial D_y} = 0. \tag{2}$$

The scalar potential $\varphi$ is found as a solution of the Poisson's equation coupled with Eqs. (1), (2):

$$\Delta\varphi = \frac{1}{\sigma}(\nabla, \mathbf{j}_{sc}), \tag{3}$$

where the superconducting current density is $\mathbf{j}_{sc} = \frac{1}{2i\kappa}(\psi^*\nabla\psi - \psi\nabla\psi^*) - \mathbf{A}|\psi|^2$ and $\sigma$ is the normal conductivity. The transport current density $j_{tr}(y) = \text{const} = j_{tr}$ is imposed via the boundary conditions for Eq. (3) at the edges of the slit, to which the electrodes are attached:

$$(\mathbf{n}, \nabla)\varphi|_{\partial D_s} = -\frac{1}{\sigma}j_{tr}; \quad (\mathbf{n}, \nabla)\varphi|_{\partial D_y} = 0. \tag{4}$$

The vector potential with components $A_s(s, y)$ and $A_y(s, y)$ (where $s \equiv R\theta$) is chosen in the Coulomb gauge

$$A_s(s, y) = 0; \quad A_y(s, y) = -BR\cos\left(\frac{s}{R}\right). \tag{5}$$

In order to guarantee the gauge invariance of the solution, we use Link Variables [28]:

$$U_s = \exp\left(-\int_{s_0}^{s} i\kappa A_s(\chi, y)d\chi\right); \quad U_y = \exp\left(-\int_{y_0}^{y} i\kappa A_y(s, \gamma)d\gamma\right). \tag{6}$$

Then the set of equations (1) to (4) takes the form:

$$\Delta\varphi = \frac{1}{2\sigma\kappa}\sum_{\mu=s,y} Im\left(\psi^* U_\mu^* \frac{\partial^2}{\partial\mu^2}[\psi U_\mu] - \psi U_\mu \frac{\partial^2}{\partial\mu^2}[\psi^* U_\mu^*]\right);$$
$$\left.\frac{\partial\varphi}{\partial s}\right|_{\partial D_s} = -\frac{1}{\sigma}j_{tr}; \quad \left.\frac{\partial\varphi}{\partial y}\right|_{\partial D_y} = 0, \quad (7)$$

$$\frac{\partial\psi}{\partial t} = \frac{1}{\kappa^2}\sum_{\mu=s,y}\left(U_\mu^* \frac{\partial^2}{\partial\mu^2}[\psi U_\mu]\right) + (1-|\psi|^2)\psi - i\kappa\varphi\psi;$$
$$\psi|_{t=0} = \Psi(s,y); \quad \left.\frac{\partial}{\partial s}[\psi U_s]\right|_{\partial D_s} = 0; \quad \left.\frac{\partial}{\partial y}[\psi U_y]\right|_{\partial D_y} = 0. \quad (8)$$

We solve the set of equations (7), (8) numerically, using the relaxation method with a random ($|\psi|$ from the range [0, 1]) initial distribution of the order parameter $\psi(s,y)$. In the presence of the magnetic field ($B > B_{c1}$) and the transport current, it evolves toward a quasi-stationary state. In the single-vortex-chain regime, the quasi-stationaly state is characterized by a periodic vortex nucleation and denucleation at the points on the edges with the highest or the lowest component of the magnetic field normal to the surface (see [18]).

Table 1. Materials and geometrical parameters used for simulation (dirty limit [2] is used)

|  | Denotation | Value (for Nb) [19] | Value (for Sn) [29-30] |
|---|---|---|---|
| Penetration depth | $\lambda = \lambda_0\sqrt{\xi_0/(2.66l(1-T/T_c))}$ | 273 nm | 269 nm [1] |
| Coherence length | $\xi = 0.855\sqrt{\xi_0 l/(1-T/T_c)}$ | 58 nm | 66 nm [2] |
| GL parameter | $\kappa = \lambda/\xi$ | 4.7 | 4.1 |
| Fermi velocity | $v_F = \sqrt{2E_F/m_e}$ | 6×10⁵ m/s | 8×10⁵ m/s [30] |
| Thickness of the film | $d$ | 50 nm | 50 nm |
| Mean free electron path | $l$ | 6.0 nm | 6.0 nm [31] |
| Diffusion coefficient | $D = lv_F/3$ | 1.2×10⁻³ m²/s | 1.6×10⁻³ m²/s |
| Relative temperature | $T/T_c$ | 0.95 | 0.77 |
| Normal conductivity for a thin membrane [19] | $\sigma = l/[3.72\times10^{-16}\ (Om\cdot m^2)]$ | 16 (μOm·m)⁻¹ | 16 (μOm·m)⁻¹ |

Table 2. Dimensionless units used for Eqs. (7) and (8)

|  | Unit | Value (Nb) | Value (Sn) |
|---|---|---|---|

---

[1] Calculated from $\lambda_0$ = 34 nm [29] for the selected relative temperature.
[2] Calculated from $\xi_0$ = 230 nm [29] for the selected relative temperature.

| Time | $\xi^2/D$ | 2.8 ps | 2.7 ps |
|---|---|---|---|
| Length | $\lambda$ | 273 nm | 269 nm |
| Magnetic field | $\Phi_0/[2\pi\lambda\xi]$ | 20.6 mT | 18.5 mT |
| Current density | $\hbar c^2/[8\pi\lambda^2\xi e]$ | 60 GA/m² | 55 GA/m² |
| Electric potential | $\sqrt{2}H_c\lambda^2/c\tau$ | 540 μV | 484 μV |
| Conductivity | $c^2/[4\pi\kappa^2 D]$ | 31 (μOm·m)⁻¹ | 30 (μOm·m)⁻¹ |

Vortices move paraxially along the tube. The moving vortices generate an electric field, which is opposite to the transport current density. We evaluate this field in terms of the voltage between the two electrodes shown in Fig. 1a $\Delta\Phi = \varphi_1 - \varphi_2$. The scalar potential $\varphi$ is found from the Poisson equation (4), for which the following spacio-temporal averaging is applied in line with the typical experimental situation (see the voltage evolution in *SI* Fig. S1):

$$\langle\Delta\Phi\rangle = \frac{1}{L_\Phi T}\int_0^T \int_0^{L_\Phi} [\varphi(s_1,y,t) - \varphi(s_2,y,t)]dydt, \qquad (9)$$

where $s_1$ and $s_2$ are $\delta/2$ and $2\pi R - \frac{\delta}{2}$, correspondingly, $\delta$ is the slit width, $L_\Phi$ is the length of the averaging area (its maximal value is the length $L$ of the tube), $T \gg \Delta t_1$ is the time of averaging, $\Delta t_1$ is the time required for a vortex to reach the opposite edge of a tube after nucleation [18].

# Figures

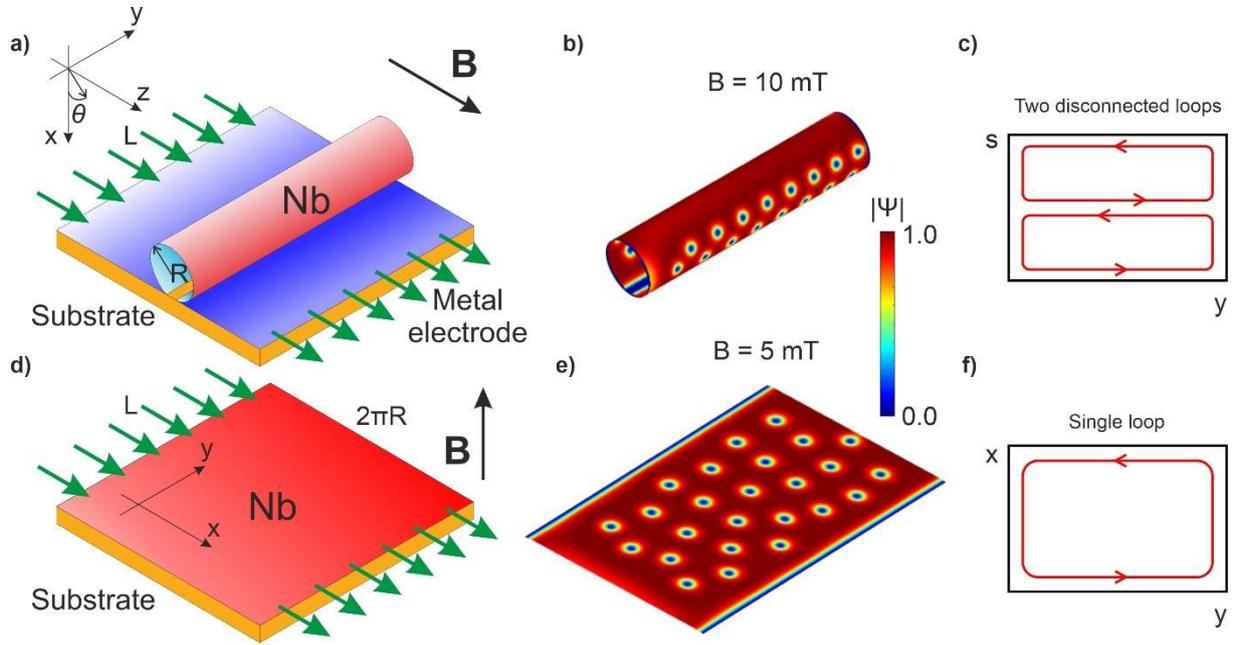

**Figure 1:** Scheme of the open tube **(a)** and the planar film **(d)**. The $y$-axis is selected as the polar axis of cylindrical coordinates $(\rho, \theta, y)$. The azimuthal angle $\theta$ is counted from the direction of the $x$-axis passing through the middle of the slit, the corresponding arc length is $s = R\theta$. The applied magnetic field $\mathbf{B} = B\mathbf{e_z}$ is directed along the $z$-axis for the tube. The transport current (green arrows) flows from the rear electrode over the cylindtical curface to the front electrode. Distribution of the modulus of the order parameter $|\psi|$ in the Nb tube of radius 500 nm **(b)** and in the corresponding unrolled planar film **(e)**. Vortices nucleate at the right edge, move in the direction opposite to the y-axis and denucleate at the left edge in the front half-tube and vice versa in the rear half-tube. Schemes of the SSC in the open tube represented on its surface **(c)** and on the planar film **(f)**.

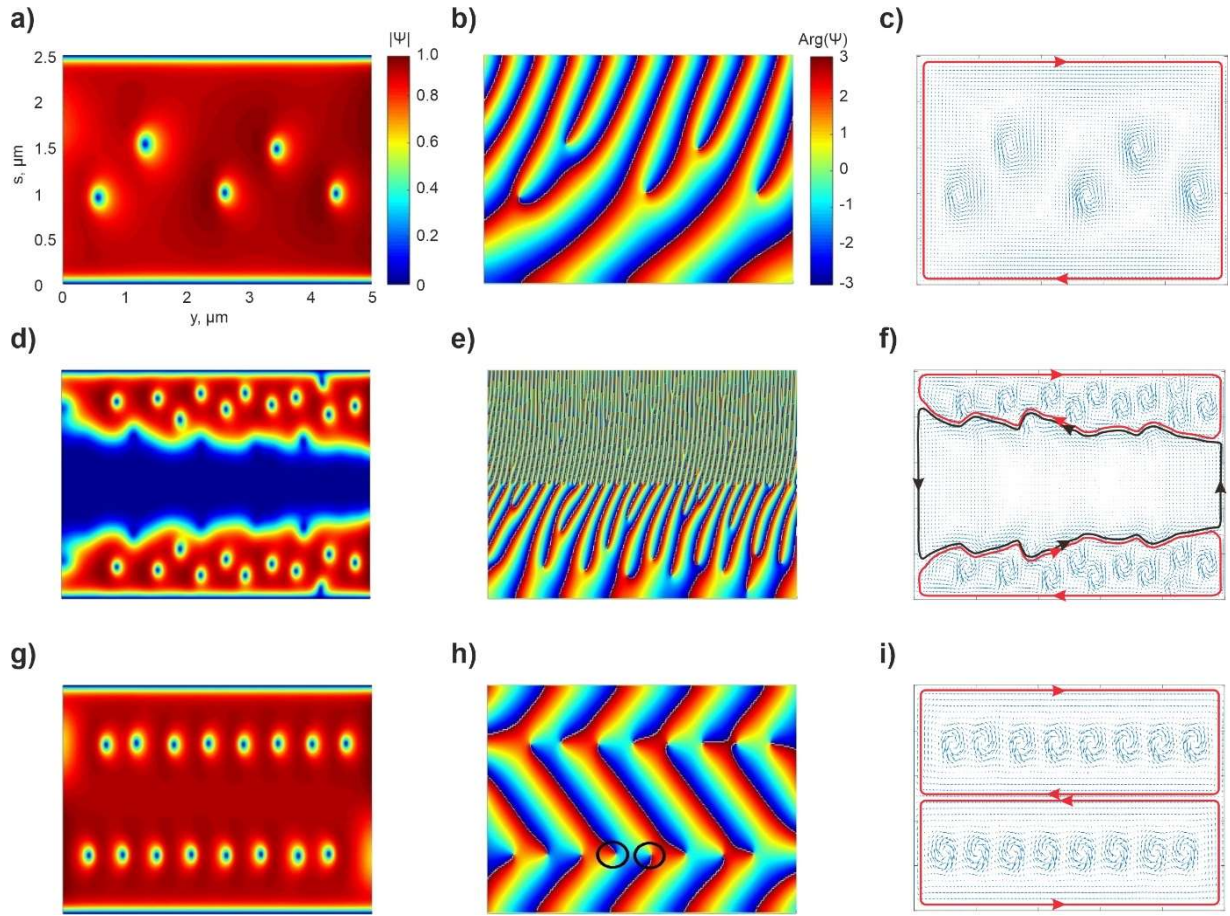

**Figure 2:** The vortex and phase-slip patterns in the planar structure under the magnetic field 2 mT **(a, b, c)** and 14 mT **(d, e, f)** and in the open tube under the magnetic field 6 mT **(g, h, i)**. In each row, the l.h.s, central and r.h.s panels represent, correspondingly, the modulus of the order parameter, the phase of the order parameter, and the current density vector field encircled by the lines of SSCs. In panel **h**, the vortex cores are marked with black circles.

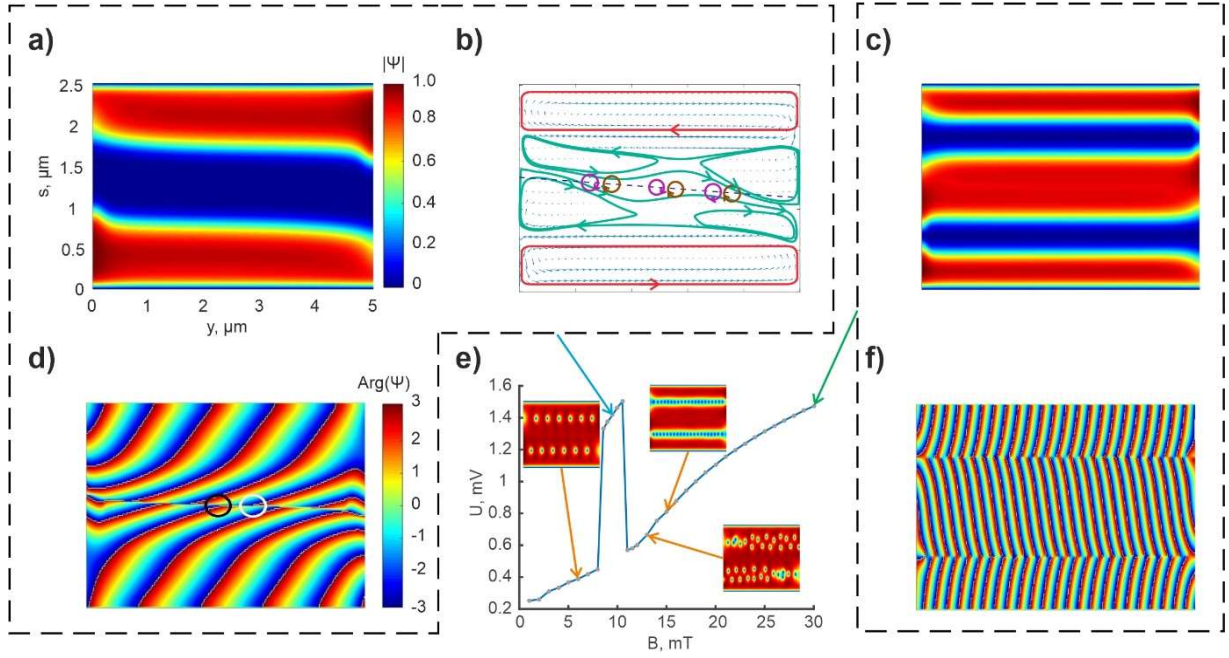

**Figure 3:** The modulus **(a)** **[(c)]** and the phase **(d)** **[(f)]** of the order parameter and the streamlines of the superconducting current **(b)** in a Nb nanotube of radius 400 nm under the applied magnetic field 10 mT [30 mT]. Red lines are disconnected loops of SSCs flowing in both half-tubes. The current distribution within the phase-slip region (blue area in panel **(a)**) consists of two new disconnected loops (green lines in **(b)**), between which the (dashed) line of fast dynamics of ***vortex-antivortex pairs*** (burgundy and brown circles) occurs. The voltage induced due to the moving vortices as a function of the magnetic field **(e)**; distributions of the modulus of the order parameter for a few magnetic fields are shown in insets. All results are obtained at $T/T_c = 0.95$ for the transport current density $j_{tr}$ = 20 GA/m$^2$.

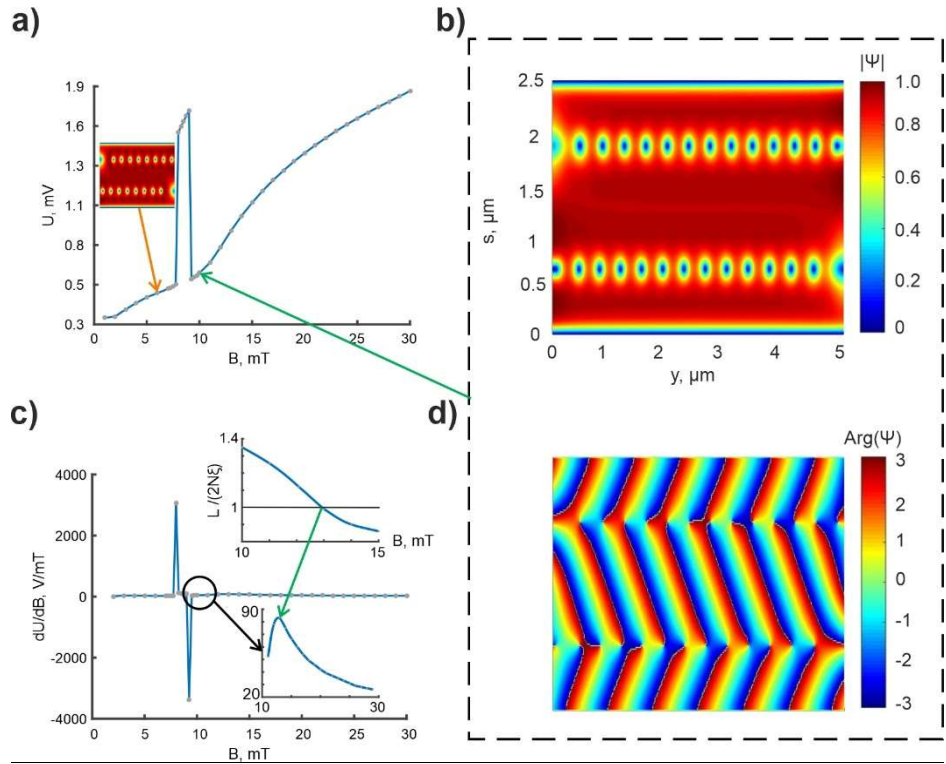

**Figure 4:** The modulus **(b)** and the phase **(d)** of the order parameter in a Sn nanotube of radius 400 nm under the applied magnetic field 9 mT. The voltage induced due to the moving vortices as a function of the magnetic field **(b)**. The voltage derivative as a function of the magnetic field **(c)**. The upper inset is the relative size of the interval per one vortex $L/(N/2)$ in a chain of $N/2$ vortices with respect to the characteristic dimension of a vortex $\sim 4\xi$. The lower inset is the zoomed plot of the voltage derivative in the magnetic field range from 10 mT to 30 mT. All results are obtained at $T/T_c = 0.77$ for the transport current density $j_{tr}$ = 17.54 GA/m$^2$.

.

**Topological transitions in superconductor nanomembranes in a magnetic field with submicron inhomogeneity under a strong transport current** (Supporting Information)

R. O. Rezaev[1,2], E. I. Smirnova[1], O. G. Schmidt[1,3-5], V. M. Fomin[1,6]

[1]*Institute for Integrative Nanosciences, Leibniz IFW Dresden, Helmholtzstraße 20, D-01069 Dresden, Germany*

[2]*Tomsk Polytechnic University, Lenin av. 30, 634050 Tomsk, Russia*

[3]*Material Systems for Nanoelectronics, TU Chemnitz, Reichenhainer Straße 70, D-09107 Chemnitz, Germany*

[4]*Research Center for Materials, Architectures and Integration of Nanomembranes (MAIN), TU Chemnitz, Rosenberg Straße 6, D-09126 Chemnitz, Germany*

[5]*Nanophysics, Faculty of Physics, TU Dresden, Nöthnitzer Straße 61, D-01187 Dresden, Germany*

[6]*National Research Nuclear University MEPhI, Kashirskoe shosse 31, 115409 Moscow, Russia*

For the vortex chain motion (Fig. S1**b**), the regular oscillations of the averaged over the contacts voltage (Fig. S1**a**) as a function of time are associated with a non-homogeneous vortex velocity. The velocity of a vortex near the nucleation point is higher as compared to its velocity in the middle or all the more so at the end of the chain near the denucleation point. This is implied by the vortex paths shown in Figs. 4a,b of Ref. [18]. Therefore, the maximal voltage induced by the moving vortices is observed, when the vortex nucleates and moves very fast, while the minimal voltage corresponds to the denucleation point, when the velocity of the vortex is the smallest.

For the phase-slip regime, the evolution of the voltage (Fig. S1**d**) dramatically differs from the vortex chain motion. The voltage asymptotically increases up to a stationary value (Fig. S1**c**).

Fig. S2 demonstrates a non-monotonic behavior of the average voltage as a function of the magnetic field at different temperatures for the Nb microtube. The phase distributions of the order parameter clearly reveal the emerging topological defects: vortices and phase slips.

The color diagram in the $B$–$J$ axes for the Sn microtube (Fig. S3**b**) shows sharp edges of the regions with high values of the voltage, which correspond to peaks when the voltage is plotted as the function of magnetic field at a specific current density value. The temperature range, where the phase-slip regime exists (Fig. S3**a**), for the Sn microtube is wider than for the Nb one (Fig. S2, **central panel**). For higher temperatures, the voltage peak becomes lower and wider.

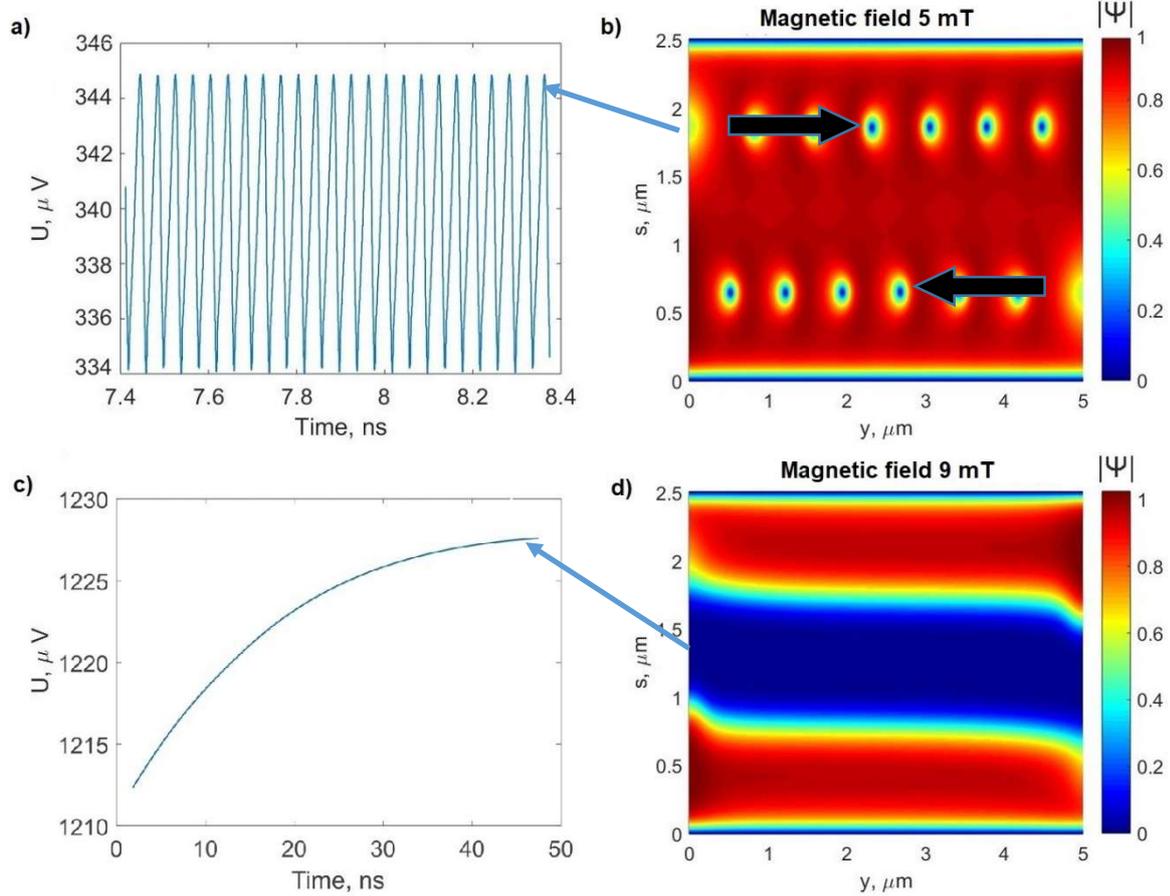

**Figure S1.** The voltage as a function of time for the vortex chain motion (**a**, **b**) and the phase slip regime (**c**, **d**) in the Sn microtube of the radius $R = 400$ nm. The temperature is $T = 0.770T_c$. The black arrows indicate the direction of the vortex chain motion. The blue arrows point to the time, at which the modulus of the order parameter is plotted. The transport current density is $j_{tr} = 17.54$ GA/m$^2$

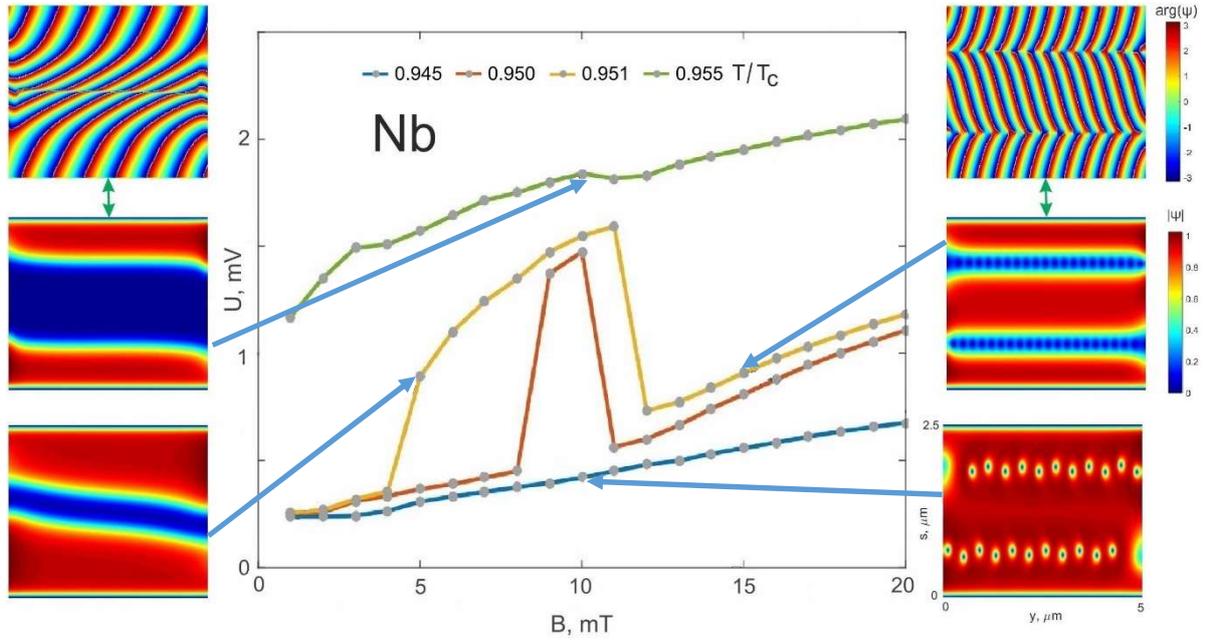

**Figure S2.** The average voltage (**central panel**) as a function of the magnetic field for the Nb microtube at different temperatures. The temperature varies from $0.945T_c$ to $0.955T_c$. The order parameter distributions are given in insets for some temperatures and magnetic fields as indicated by blue arrows. Green arrows show the correspondence of the phase and modulus distributions. The transport current density is $j_{tr} = 20.0$ GA/m$^2$.

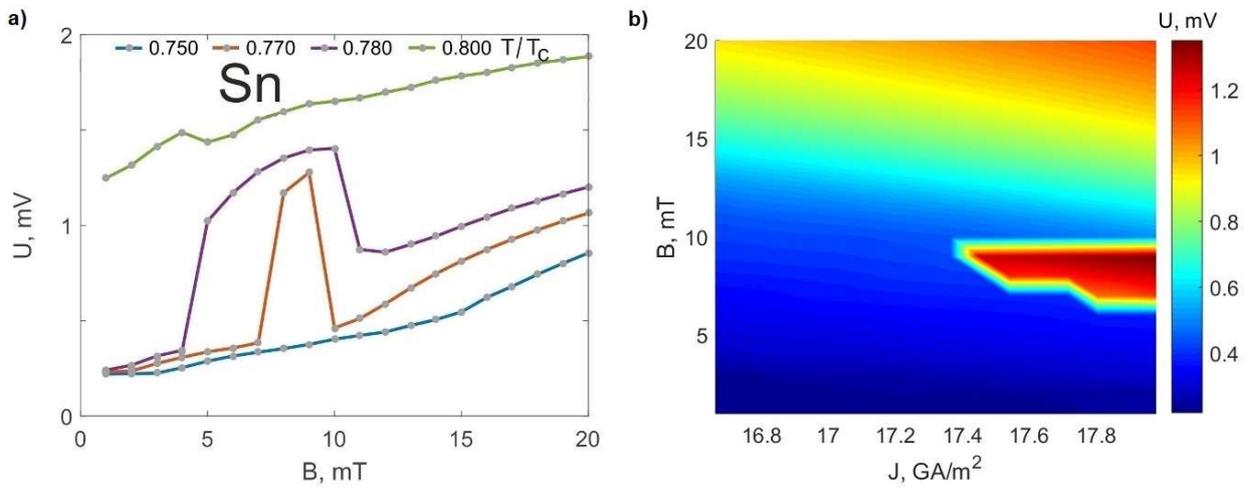

**Figure S3.** The average voltage (**a**) as a function of the magnetic field for the Sn microtube at different temperatures. The temperature varies from $0.75\ T_c$ to $0.8T_c$. The voltage distribution is plotted as a color diagram in the $B$–$J$ axes (**b**) at the temperature $T=0.77T_c$. The transport current density is $j_{tr} = 17.54$ GA/m$^2$.